# From Rigid to Adiabatic: Canonical Regularization of AC Networks via Action-Angle Variables

Feng Ji, Lu Gao, and Lihui Yang

***Abstract*—Traditional power system analysis relies on timescale separation and the rigid-network assumption, freezing electromagnetic transients into algebraic power-flow equations via Steinmetz's phasor theory. As grid-forming converter penetration increases, magnetic energy dynamics on transmission lines interact with converter control loops on comparable timescales, challenging this rigid-network assumption. Returning to Faraday's law of electromagnetic induction driven by rotating magnetic fields, this paper models the transmission lines' rotating magnetic fields in action-angle canonical coordinates, regularizes the rigid algebraic constraints of power-flow equations into canonical equations on adiabatic symplectic manifolds, and establishes a port-Hamiltonian standard form for AC power grids. Based on the minimal-counterexample principle and using the equal-area criterion's classical two-machine system, this paper reveals a latitudinal instability channel via Bloch-sphere coordinates: *Q-V* control releases voltage-amplitude freedom, shifting the stability boundary from the equatorial UEP(unstable equilibrium point) to a saddle point, thereby unifying the analytical frameworks of *P-δ* angle stability and *Q-V* voltage stability in power system analysis.**

## I. INTRODUCTION

Traditional power system analysis relies on the assumption of timescale separation: through the rigidity assumption, the electromagnetic transient processes of the network are compressed into algebraic constraints that hold instantaneously—namely, algebraic power-flow equations—while the dynamics of generators, inverters, and other apparatus are described by differential equations. This DAE (Differential Algebraic Equation) framework has a solid physical foundation in the era of bulk power systems dominated by synchronous machines, because the electromechanical transient timescale is orders of magnitude larger than the electromagnetic transient timescale. In analyzing electromechanical dynamics on the timescale of seconds, the grid dynamics on the millisecond timescale can be safely frozen as rigid algebraic snapshots[1]-[3]. On this premise, energy-function methods such as the direct method [4], PEBS (Potential-Energy Boundary Surface)[5] and BCU (Boundary of stability region based on Controlling Unstable equilibrium point)[6] were developed, providing direct criteria for transient-stability analysis.

However, with the increasing penetration of grid-forming converters, the magnetic energy dynamics on transmission lines and the dynamics of inverter control loops lie on comparable timescales, posing a challenge to the traditional modeling approach based on timescale separation. Traditional algebraic power-flow equations assume that the grid is rigid and that electromagnetic transients are much faster than electromechanical transients[1]; yet, under high penetration of power-electronic-based devices, this assumption faces challenges such as timescale coupling[7]. Reference [8] points out that simplified models based on quasi-steady-state impedance equivalents may fail to accurately capture key dynamics during transient processes, and can even lead to overly conservative stability conditions. Reference [9] reviews the positive-feedback mechanism of the reactive-power control loop of grid-forming converters after faults and its coupling-induced deterioration effect on angle dynamics, highlighting the constraint of critical voltage magnitude on stability. Nevertheless, these analyses are scattered across three mutually incompatible frameworks—the *P-δ* plane, the *Q-V* plane, and time-domain simulation[9]—lacking a unified energy criterion and stability-region analysis theory.

PHS (Port-Hamiltonian system) theory is a branch of modern control theory which provides a unified Hamiltonian energy analysis and control framework. It has been widely applied to energy shaping, passivity-based control, and stability analysis of inverters and synchronous machines[10][11]-[12], demonstrating unique advantages in modeling and control of power apparatus. However, existing literature treats the AC networks as rigid algebraic equations, thereby converting them into algebraic constraints of static ports in the Dirac structure of the PHS framework[11][12]. To incorporate network transients into the unified energy framework, dynamic canonical equations must be established by first identifying the Hamiltonian and the canonical coordinates, so as to restore the network constraints to autonomous flows driven by energy gradients. Only via canonical regularization can PHS be rigorously applied to AC networks, thereby unifying the energy language of both apparatus and networks.

In conventional circuit analysis, inductor current and capacitor voltage form a natural canonical pair. Accordingly, references [8], [13] and [14] employ PI-equivalent circuits for transmission line modeling, reintroducing equivalent shunt capacitances that traditional electromechanical transient analysis neglects. The voltage of the shunt capacitance and the

F. Ji and L. Gao are with State Key Laboratory of Advanced Transmission Technology (China Electric Power Research Institute Ltd), Changping District, Beijing 102200, China. L. Gao and L. Yang (Corresponding author) are with School of Electrical Engineering, Xi'an Jiaotong University, Xi'an 710049, China (e-mail: jameskeating@163.com； lihui.yang@xjtu.edu.cn).

current of the series inductance thereby constitute canonical coordinate pairs, achieving regularization of the AC networks. However, this reintroduction of shunt capacitances causes the model dimension to expand, progressively departing from the compactness of power-flow equations.

Modern physics routinely employs action-angle variables as canonical coordinates to model rotating systems mathematically[15]-[17]. Action is defined as the ratio of energy to angular velocity, so it physically corresponds to the angular momentum of the system. Since the energy is independent of (or cyclic in) the angle variable, Noether's theorem directly yields the invariance of its conjugate action[16]. When system parameters vary slowly, that is the adiabatic process, this conserved quantity further becomes an adiabatic invariant[16][17]. It is precisely by virtue of closed-form integrability, explicit conservation laws, and adiabatic robustness that action-angle variables have become the standard language for describing rotating systems in celestial mechanics[18][19], quantum mechanics[20], and plasma physics[21].

AC power transmission is rooted in Faraday's law of electromagnetic induction driven by rotating magnetic fields; therefore, the action-angle variable framework can be adopted to embed the power-flow algebraic equations into the canonically regularized port-Hamiltonian system theory, thereby obtaining energy landscapes and stability boundaries. Following the argumentation tradition of the "minimal counterexample" in theoretical physics, this paper takes the classical two-node system, upon which the EAC (equal-area criterion) is founded, as the core research subject, and conducts a rigorous comparative study between the proposed regularized energy landscape and the classical equal-area criterion. The hierarchy of AC network modeling frameworks in this paper is shown in Table I and the main contributions include:

(1) By adopting the magnetic energy as the Hamiltonian energy and selecting nodal action and angle variables as the canonical coordinates, this paper establishes the standard PHS for AC networks;

(2) This paper discovers the conservation law of the total nodal action during adiabatic processes, revealing the causal chain in which reactive-power imbalance directly drives angle evolution;

(3) By introducing the Bloch-sphere coordinates for energy-landscape visualization, this paper reveals the saddle points on the energy landscape, and reveals the saddle-point-induced instability triggered by reactive-power voltage control.

The remainder of this paper is organized as follows: Section II establishes the basic concepts of action-angle variables using a single-node grounded inductor as an example. Section III derives the power expressions and the regularized Hamiltonian for a two-node model, and establishes its standard PHS form. Section IV uses Bloch-sphere coordinates to visualize the energy landscape. Section V discusses the modification of the energy landscape under *P*-$\delta$ control. Section VI discusses the modification of the energy landscape under *Q*-*V* control. Section VII presents the action-angle regularized model for multi-node multi-branch power networks. Finally, Section VIII concludes the paper.

TABLE I
Hierarchy of AC Network Modeling Frameworks

| Framework | EMT model | Adiabatic model (This work) | Algebraic model |
|---|---|---|---|
| Line Model | PI-equivalent circuits | Inductor | Impedance |
| Governing Equation | $dx/dt=f(x,u)$ | $dx/dt=J\nabla H(x)+g(x)u$ | $0=g(x,u)$ |
| Node Variable | $u_a, u_b, u_c,$ $i_a, i_b, i_c$ | $\mathcal{J}, \theta$ | $u_{\text{real}}, u_{\text{imag}}$ |
| Timescale | µs~ms | ms~s | s~min |

Note: TABLE I summarizes the hierarchy, EMT resolves the full $\pi$-section dynamics on the microsecond scale; the proposed adiabatic model operates on the 10-ms step, the same as traditional electromechanical transient DAE equations, yet it explicitly restores the dynamic evolution of network magnetic energy.

## II Regularization of A Single-Node System

**Definition:** In Hamiltonian mechanics, a pair of variables $(q, p)$ is said to be canonical if there exists a scalar function $H(q, p)$ for which the system dynamics satisfy the canonical equations (1).

$$\dot{q} = \partial H/\partial p,\ \dot{p} = -\partial H/\partial q \tag{1}$$

where $H$ is referred to as the Hamiltonian (or energy function).

In classical mechanics textbooks, the canonical action-angle variable framework is usually introduced using the one-dimensional harmonic oscillator as the standard pedagogical example [16][17], because the action integral of its periodic motion admits a closed-form evaluation. For more general rotating systems such as rigid bodies, the conjugate momenta expressed in Euler angles involve complicated elliptic integrals; therefore, this framework is usually avoided in standard introductory textbooks. However, the physical structure of rotating magnetic fields in AC systems—which reduces to uniform circular motion in the synchronous rotating reference frame—enables the construction of action-angle variables without resorting to elliptic integrals, thereby allowing this framework to be systematically applied to rotating electromagnetic systems.

In AC circuit analysis, for a node connected to ground potential through a three-phase inductor *L*, the energy in the inductor can be calculated as

$$H = \frac{\Psi_\alpha^2 + \Psi_\beta^2}{2L} \tag{2}$$

where $\Psi_\alpha$ and $\Psi_\beta$ are the two projected components of the nodal flux linkage $\Psi$ in the rotating $\alpha\beta$ coordinates. Since the other node of the inductor is grounded, the nodal flux linkage $(\Psi_\alpha, \Psi_\beta)$ is also the branch flux linkage. The concept of nodal flux linkage as a Lagrangian coordinate for AC networks is systematically established in [22].

We select the *xy* coordinate system rotating with uniform angular velocity $\omega$. References [22] and [23] give the correspondence between the stationary $\alpha\beta$ coordinate system and the rotating *xy* coordinate system, as expressed in (3). Additionally, the differential relationship between the nodal flux linkage and the nodal voltage in the *xy* coordinate system is given by (4).

$$\begin{bmatrix} \Psi_\alpha \\ \Psi_\beta \end{bmatrix} = \begin{bmatrix} \cos(\omega t) & -\sin(\omega t) \\ \sin(\omega t) & \cos(\omega t) \end{bmatrix} \begin{bmatrix} \Psi_x \\ \Psi_y \end{bmatrix} \tag{3}$$

$$\begin{bmatrix} \dot{\Psi}_x \\ \dot{\Psi}_y \end{bmatrix} = \begin{bmatrix} 0 & \omega \\ -\omega & 0 \end{bmatrix} \begin{bmatrix} \Psi_x \\ \Psi_y \end{bmatrix} + \begin{bmatrix} u_x \\ u_y \end{bmatrix} \tag{4}$$

Alternatively, using complex notation, (3) can be written as

$$\Psi_{\alpha\beta} = e^{j\omega t}\Psi \tag{5}$$

where $\Psi_{\alpha\beta} = \Psi_\alpha + j\Psi_\beta$ and $\Psi = \Psi_x + j\Psi_y$.

Equation (4) can be written as

$$\dot{\Psi} = -j\omega\Psi + u\,,\ u = u_x + ju_y \tag{6}$$

Taking the coordinate transformation $\mathcal{J} = |\Psi|^2 / (2\omega L)$ and $\theta = \text{angle}(\Psi)$, the inductor energy function (2) can be written as

$$H = \omega\mathcal{J} \tag{7}$$

If the input $u = 0$, we have (8).

$$\begin{cases} \dot{\mathcal{J}} = \dfrac{\partial H}{\partial \theta} = 0 \\ \dot{\theta} = -\dfrac{\partial H}{\partial \mathcal{J}} = -\omega \end{cases} \tag{8}$$

Equation (8) satisfies the canonical regularization requirements prescribed in (1). Hence, $(\mathcal{J}, \theta)$ constitutes a canonical pair. $\mathcal{J} = H/\omega$ is the nodal action, equal to the area enclosed by the nodal flux linkage trajectory in the $\alpha\beta$ plane divided by 2π.

The negative sign in $\dot{\theta} = -\omega$ in (8) reflects the definition of $\theta = \text{angle}(\Psi_x + j\Psi_y)$ as the argument of the flux linkage in the rotating frame. With no external input ($u = 0$), the flux linkage vector is stationary in the $\alpha\beta$ frame, equivalent to a uniform $-\omega$ rotation in the *xy* frame.

If the input $u \neq 0$, it can be shown that, in the $(\mathcal{J}, \theta)$ coordinates, (4) transforms into (9) (see Appendix A for the detailed derivation).

$$\begin{cases} \dot{\mathcal{J}} = 0 + \sqrt{\dfrac{2\mathcal{J}}{\omega L}}\left(u_x \cos\theta + u_y \sin\theta\right) \\ \dot{\theta} = -\omega + \dfrac{-u_x \sin\theta + u_y \cos\theta}{\sqrt{2\omega L \mathcal{J}}} \end{cases} \tag{9}$$

From (9), it is evident that $\dot{\mathcal{J}}$ depends only on the projection of the voltage onto the flux-linkage direction (the radial component), while $\dot{\theta}$ depends only on the projection of the voltage perpendicular to the flux-linkage direction (the tangential component). Alternatively, (9) can be written in the standard PHS form (10).

$$\dot{x} = J\nabla H(x) + g(x)u \tag{10}$$

where $x = \begin{bmatrix} \mathcal{J} \\ \theta \end{bmatrix}$, $u = \begin{bmatrix} u_x \\ u_y \end{bmatrix}$, $J = \begin{bmatrix} 0 & 1 \\ -1 & 0 \end{bmatrix}$,

$\nabla H = \begin{bmatrix} \dfrac{\partial H}{\partial \mathcal{J}} \\ \dfrac{\partial H}{\partial \theta} \end{bmatrix} = \begin{bmatrix} \omega \\ 0 \end{bmatrix}$, $g(x) = \begin{bmatrix} \gamma\cos\theta & \gamma\sin\theta \\ -\eta\sin\theta & \eta\cos\theta \end{bmatrix}$, $\gamma = \sqrt{\dfrac{2\mathcal{J}}{\omega L}}$,

$\eta = \dfrac{1}{\sqrt{2\omega L\mathcal{J}}} = \dfrac{\gamma}{2\mathcal{J}}$.

## III Regularization of The Two-Node System

This paper considers the classical two-node system shown in Fig. 1, which is commonly adopted in the equal-area criterion, and unifies *P*-*δ* and *Q*-*V* stability analysis within the regularized action-angle framework. In Fig. 1, the rated three-phase voltage is 690 V, the frequency is 50 Hz, and the line inductance is *L*=0.5mH. The rated transmission capacity is 3 MW, corresponding to a short-circuit ratio SCR=1.01.

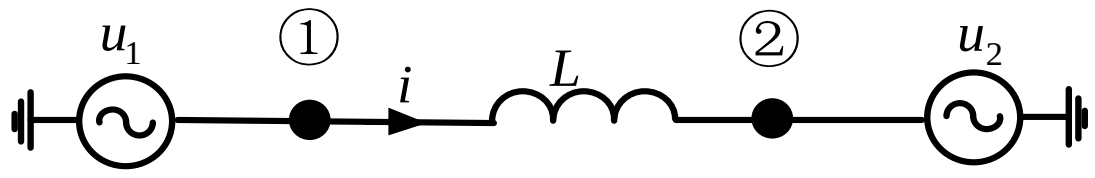

**Fig. 1.** Two node single branch AC system.

The mathematical model of the system in Fig. 1 is given by (11). When the dynamic processes are neglected, (11) reduces to the conventional power-flow equations (see Appendix B for the detailed proof).

$$\frac{d}{dt}\begin{bmatrix} \Psi_{1x} \\ \Psi_{1y} \\ \Psi_{2x} \\ \Psi_{2y} \end{bmatrix} = \begin{bmatrix} 0 & \omega & 0 & 0 \\ -\omega & 0 & 0 & 0 \\ 0 & 0 & 0 & \omega \\ 0 & 0 & -\omega & 0 \end{bmatrix}\begin{bmatrix} \Psi_{1x} \\ \Psi_{1y} \\ \Psi_{2x} \\ \Psi_{2y} \end{bmatrix} + \begin{bmatrix} u_{1x} \\ u_{1y} \\ u_{2x} \\ u_{2y} \end{bmatrix} \tag{11}$$

Or simply denoted as

$$\dot{\Psi}_k = -j\omega\Psi_k + u_k\,,\ k = 1, 2 \tag{12}$$

Taking the power injected at node 1 as an example, based on the instantaneous power theory[27], we have

$$\begin{aligned} S_1 = P_1 + jQ_1 = u_1 \cdot i^* &= \left(\dot{\Psi}_1 + j\omega\Psi_1\right)\cdot\left(\frac{\Psi_1 - \Psi_2}{L}\right)^* \\ &= \frac{1}{L}\left[\dot{\Psi}_1\Psi_1^* - \dot{\Psi}_1\Psi_2^* + j\omega|\Psi_1|^2 - j\omega\Psi_1\Psi_2^*\right] \end{aligned} \tag{13}$$

where the superscript $*$ denotes complex conjugation.

The active power is the real part of (13). The real part of each of the four terms is listed as Table II.

Therefore, the complete active-power expression is

$$P_1 = \frac{1}{L}\left[\omega|\Psi_1||\Psi_2|\sin(\theta_1 - \theta_2) + \frac{1}{2}\frac{\mathrm{d}}{\mathrm{dt}}|\Psi_1|^2 - \dot{\Psi}_1 \cdot \Psi_2\right] \tag{14}$$

Obviously, in steady state, all differential terms in (14) vanish, and it reduces to the classical power-angle equation (15):

$$P_{1_steady} = \frac{\omega|\Psi_1||\Psi_2|\sin(\theta_1-\theta_2)}{L} = \frac{|u_1||u_2|\sin(\theta_1-\theta_2)}{\omega L} \quad (15)$$

The reactive power is obtained by taking the imaginary part of (13), with the four terms listed in Table III.

TABLE II
DESCRIPTION OF ACTIVE-POWER TERMS

| Term | Real-Part | Physical Meaning |
|---|---|---|
| $\dot{\Psi}_1\Psi_1^*$ | $\dot{\Psi}_{1x}\Psi_{1x}+\dot{\Psi}_{1y}\Psi_{1y} = \frac{1}{2}\frac{\mathrm{d}}{\mathrm{dt}}\|\Psi_1\|^2$ | Rate of change of local magnetic-energy |
| $-\dot{\Psi}_1\Psi_2^*$ | $-(\dot{\Psi}_{1x}\Psi_{2x}+\dot{\Psi}_{1y}\Psi_{2y}) = -\dot{\Psi}_1\cdot\Psi_2$ | Dynamic flux-linkage coupling between the two nodes |
| $j\omega\|\Psi_1\|^2$ | 0 (purely imaginary) | No contribution |
| $-j\omega\Psi_1\Psi_2^*$ | $\omega\|\Psi_1\|\|\Psi_2\|\sin(\theta_1-\theta_2)$ | Steady-state synchronous power (classical term) |

TABLE III
DESCRIPTION OF REACTIVE-POWER TERMS

| Term | Imaginary-Part | Physical Meaning |
|---|---|---|
| $\dot{\Psi}_1\Psi_1^*$ | $\dot{\Psi}_{1y}\Psi_{1x}-\dot{\Psi}_{1x}\Psi_{1y}=\|\Psi_1\|^2\dot{\theta}_1$ | Local reactive power due to flux-linkage rotation |
| $-\dot{\Psi}_1\Psi_2^*$ | $-(\dot{\Psi}_{1y}\Psi_{2x}-\dot{\Psi}_{1x}\Psi_{2y})$ | Rotational flux-linkage coupling between the two nodes |
| $j\omega\|\Psi_1\|^2$ | $\omega\|\Psi_1\|^2$ | Local reactive-power reference term |
| $-j\omega\Psi_1\Psi_2^*$ | $-\omega\|\Psi_1\|\|\Psi_2\|\cos(\theta_1-\theta_2)$ | Steady-state coupling reactive power |

Therefore, the complete reactive-power expression is

$$Q_1 = \frac{1}{L}\begin{bmatrix} \omega|\Psi_1|^2 - \omega|\Psi_1||\Psi_2|\cos(\theta_1-\theta_2) \\ +\dot{\Psi}_{1y}\cdot(\Psi_{1x}-\Psi_{2x}) - \dot{\Psi}_{1x}\cdot(\Psi_{1y}-\Psi_{2y}) \end{bmatrix} \quad (16)$$

Considering the current-component relationships $\Psi_{1x}-\Psi_{2x}=Li_x$ and $\Psi_{1y}-\Psi_{2y}=Li_y$, (16) can be simplified to

$$Q_1 = \frac{1}{L}\begin{bmatrix} \omega|\Psi_1|^2 - \omega|\Psi_1||\Psi_2|\cos(\theta_1-\theta_2) \\ +L\dot{\Psi}_{1y}i_x - L\dot{\Psi}_{1x}i_y \end{bmatrix} \quad (17)$$

In steady state, the reactive-power equation reduces to the classical power-angle equation, as follows:

$$Q_{1_\mathrm{steady}} = \frac{1}{L}\left[\omega\|\Psi_1\|^2 - \omega\|\Psi_1\|\|\Psi_2\|\cos(\theta_1-\theta_2)\right] \quad (18)$$

The energy function on the line inductor can be expressed as the identity shown in (19), whose derivation uses the cosine formula for the difference of vector magnitudes:

$$\begin{aligned} H &= \frac{1}{2L}|\Psi_1-\Psi_2|^2 \\ &= \frac{1}{2L}\left(|\Psi_1|^2+|\Psi_2|^2-2|\Psi_1||\Psi_2|\cos(\theta_1-\theta_2)\right) \end{aligned} \quad (19)$$

Adopting the action-angle variable framework, (19) is rewritten as

$$H = \omega\mathcal{J}_1 + \omega\mathcal{J}_2 - 2\omega\sqrt{\mathcal{J}_1}\sqrt{\mathcal{J}_2}\cos(\theta_1-\theta_2) \quad (20)$$

where $\mathcal{J}_k = \frac{|\Psi_k|^2}{2\omega L}, \theta_k = \mathrm{angle}(\Psi_k), k=1,2$

Take (20) as the Hamiltonian, we need to verify whether $(\mathcal{J}_k,\theta_k)$ satisfy the canonical equations (21).

$$\dot{\mathcal{J}}_k = \frac{\partial H}{\partial\theta_k}, \dot{\theta}_k = -\frac{\partial H}{\partial\mathcal{J}_k} \quad (21)$$

Expanding the first equation in (21), we have

$$\begin{cases} \dot{\mathcal{J}}_1 = 2\omega\sqrt{\mathcal{J}_1\mathcal{J}_2}\sin(\theta_1-\theta_2) = P_{1_steady} \\ \dot{\mathcal{J}}_2 = -2\omega\sqrt{\mathcal{J}_1\mathcal{J}_2}\sin(\theta_1-\theta_2) = P_{2_steady} \end{cases} \quad (22)$$

From (22), we obtain the conservation law of the total action $\mathcal{J}_{\mathrm{sum}} = \mathcal{J}_1+\mathcal{J}_2 = \mathrm{const}$, i.e. $\dot{\mathcal{J}}_{\mathrm{sum}} = \dot{\mathcal{J}}_1+\dot{\mathcal{J}}_2 = 0$.

Expanding the second equation in (21), we have

$$\begin{cases} \dot{\theta}_1 = -\frac{\partial H}{\partial\mathcal{J}_1} = -\frac{\omega}{\sqrt{\mathcal{J}_1}}\left(\sqrt{\mathcal{J}_1}-\sqrt{\mathcal{J}_2}\cos(\theta_1-\theta_2)\right) \\ \dot{\theta}_2 = -\frac{\partial H}{\partial\mathcal{J}_2} = -\frac{\omega}{\sqrt{\mathcal{J}_2}}\left(\sqrt{\mathcal{J}_2}-\sqrt{\mathcal{J}_1}\cos(\theta_1-\theta_2)\right) \end{cases} \quad (23)$$

Multiplying both sides of the first equation in (23) by $2\omega\mathcal{J}_1$, and considering with (18), we have

$$\begin{aligned} 2\omega\mathcal{J}_1\dot{\theta}_1 &= -2\omega^2\left(\mathcal{J}_1-\sqrt{\mathcal{J}_1\mathcal{J}_2}\cos(\theta_1-\theta_2)\right) \\ &= -\frac{1}{L}\left[\omega|\Psi_1|^2-\omega|\Psi_1||\Psi_2|\cos(\theta_1-\theta_2)\right] \\ &= -Q_{1_\mathrm{steady}} \end{aligned} \quad (24)$$

Then we obtain $\dot{\theta}_1 = \frac{-Q_{1_\mathrm{steady}}}{2\omega\mathcal{J}_1}$. By defining the relative angle as $\delta = \theta_1-\theta_2$, we have

$$\dot{\delta} = \dot{\theta}_1-\dot{\theta}_2 = \frac{-Q_{1_\mathrm{steady}}}{2\omega\mathcal{J}_1} + \frac{Q_{2_\mathrm{steady}}}{2\omega\mathcal{J}_2} \quad (25)$$

Equation (25) reveals that the change rate of the relative angle $\dot{\delta}$, equals the difference between the reactive powers per unit action $-Q_{1_\mathrm{steady}}/(2\omega\mathcal{J}_1)+Q_{2_\mathrm{steady}}/(2\omega\mathcal{J}_2)$. That is, the reactive power in the line drives the variation of the angular variable. Equation (25) can be further written as

$$\dot{\delta} = -\frac{\partial H}{\partial\mathcal{J}_1} + \frac{\partial H}{\partial\mathcal{J}_2} = -\frac{\partial H}{\partial(\mathcal{J}_1-\mathcal{J}_2)} \quad (26)$$

This indicates that the symmetries brought by the Hamiltonian canonical equations in the action-angle variable framework render the conservation laws predicted by Noether's theorem (the conservation of action, adiabatic invariance) manifest, as summarized in Table IV.

Reactive-power balance requires the relative rotation of the two nodal flux linkages to be synchronized, i.e. $\dot{\delta}=0$. It does not require each individual node to be synchronized with the *xy* coordinate system. Even if the two nodes are jointly driven by external sources to rotate at a non-free angular velocity, reactive-power balance is achieved as long as the relative angular velocity is zero.

TABLE IV
CONSERVATION LAWS CORRESPONDING TO POWER BALANCE

| Physical Meaning | Translational symmetry | Conserved Quantity |
|---|---|---|
| Active-power balance | relative angle $\delta$ | total action $\mathcal{J}_1+\mathcal{J}_2$ |
| Reactive-power balance | relative action $\mathcal{J}_1-\mathcal{J}_2$ | relative angle $\delta$ |

Based on (22) and (23), and following the same approach as in (9), the external voltage input at each node is incorporated to obtain the complete fourth-order equations (27).

$$\begin{cases}\dot{\mathcal{J}}_1=2\omega\sqrt{\mathcal{J}_1\mathcal{J}_2}\sin(\theta_1-\theta_2)+\sqrt{\dfrac{2\mathcal{J}_1}{\omega L}}\left(u_{1x}\cos\theta_1+u_{1y}\sin\theta_1\right)\\ \dot{\theta}_1=-\dfrac{\omega}{\sqrt{\mathcal{J}_1}}\left(\sqrt{\mathcal{J}_1}-\sqrt{\mathcal{J}_2}\cos(\theta_1-\theta_2)\right)+\dfrac{-u_{1x}\sin\theta_1+u_{1y}\cos\theta_1}{\sqrt{2\omega L\mathcal{J}_1}}\\ \dot{\mathcal{J}}_2=-2\omega\sqrt{\mathcal{J}_1\mathcal{J}_2}\sin(\theta_1-\theta_2)+\sqrt{\dfrac{2\mathcal{J}_2}{\omega L}}\left(u_{2x}\cos\theta_2+u_{2y}\sin\theta_2\right)\\ \dot{\theta}_2=-\dfrac{\omega}{\sqrt{\mathcal{J}_2}}\left(\sqrt{\mathcal{J}_2}-\sqrt{\mathcal{J}_1}\cos(\theta_1-\theta_2)\right)+\dfrac{-u_{2x}\sin\theta_2+u_{2y}\cos\theta_2}{\sqrt{2\omega L\mathcal{J}_2}}\end{cases}\tag{27}$$

Or, written in the standard PHS form (28).

$$\dot{x}=J\nabla H(x)+g(x)u\tag{28}$$

where $x=\begin{bmatrix}x_1\\x_2\end{bmatrix}$, $x_i=\begin{bmatrix}\mathcal{J}_i\\\theta_i\end{bmatrix}$, $u=\begin{bmatrix}u_1\\u_2\end{bmatrix}$, $u_k=\begin{bmatrix}u_{ix}\\u_{iy}\end{bmatrix}$,

$$\nabla H(x)=\begin{bmatrix}Q_{1_steady}/(2\mathcal{J}_1\omega)\\P_{1_steady}/\omega\\Q_{2_steady}/(2\mathcal{J}_2\omega)\\P_{2_steady}/\omega\end{bmatrix},\ J=\begin{bmatrix}J_1&0\\0&J_2\end{bmatrix},\ J_i=\begin{bmatrix}0&1\\-1&0\end{bmatrix},$$

$$g=\begin{bmatrix}g_1&0\\0&g_2\end{bmatrix},\ g_i=\begin{bmatrix}\gamma_i\cos\theta_i&\gamma_i\sin\theta_i\\-\eta_i\sin\theta_i&\eta_i\cos\theta_i\end{bmatrix},\ \gamma_i=\sqrt{\frac{2\mathcal{J}_i}{\omega L}},$$

$$\eta_i=\frac{1}{\sqrt{2\omega L\mathcal{J}_i}}=\frac{\gamma_i}{2\mathcal{J}_i},\ i=1,2.$$

The derivation in this section reveals a causal mechanism obscured by traditional algebraic power-flow equations: under the regularized framework, the direct driving force for angle variation does not originate from active power itself, but from reactive-power imbalance. This should not be confused with the rotor swing equation; rather, it reveals how reactive-power imbalance governs the phase dynamics of the electromagnetic state in the regularized Hamiltonian framework.

Specifically, when an active-power disturbance occurs in the system, it forces a redistribution of reactive power through power coupling, thereby breaking the reactive-power balance condition during the transient process; this reactive-power imbalance directly drives the angle to change until the system re-establishes reactive-power balance at a new steady-state operating point. In other words, there exists an indirect driving chain in the transient process—active-power change induces reactive-power imbalance, ultimately leading to angle drift. This causal chain is not captured in traditional algebraic power-flow equations.

According to the classical adiabatic theorem[16][17], the total action $\mathcal{J}_{\text{sum}}=\mathcal{J}_1+\mathcal{J}_2$ remains an adiabatic invariant during slow voltage regulation transients. Since $\mathcal{J}_{\text{sum}}\propto|\Psi_1|^2+|\Psi_2|^2\propto|u_1|^2+|u_2|^2$, conserving the total action is equivalent to holding the mean squared nodal-voltage magnitude $\frac{1}{2}\left(|u_1|^2+|u_2|^2\right)$ of the transmission line invariant. It carries transparent physical meaning and immediate engineering relevance. From a control perspective, preserving this invariant naturally emerges as the energy-level coordination objective for grid-forming converter clusters, rather than a local voltage-regulation task at individual nodes.

This conservation law constrains the system to an equi-action surface, enabling the geometric reduction to the Bloch sphere $S^2$ introduced in the next section.

## IV VISUALIZATION ON BLOCH-SPHERE

The Bloch sphere is a two-dimensional unit sphere $S^2$, which was always used to represent the state space of two-level quantum systems[24]-[26]. Every point $\boldsymbol{r}=(X,Y,Z)$ on the unit sphere uniquely corresponds to a unique system state, and the three coordinates satisfy the normalization constraint $X^2+Y^2+Z^2=1$. This geometric structure transforms abstract system-state analysis into visual spherical-geometry analysis.

The energy function given by (20) is rewritten as

$$H(\mathcal{J}_1,\mathcal{J}_2,\delta)=\omega\mathcal{J}_1+\omega\mathcal{J}_2-2\omega\sqrt{\mathcal{J}_1}\sqrt{\mathcal{J}_2}\cos(\delta)\tag{29}$$

The energy function $H(\mathcal{J}_1,\mathcal{J}_2,\delta)$ has three degrees of freedom; however, because the total action $\mathcal{J}_{\text{sum}}=\mathcal{J}_1+\mathcal{J}_2$ is conserved during the adiabatic process, the system is actually constrained on a constant-action surface. Introducing the Bloch-sphere coordinate transformation (30), we can directly verify $X^2+Y^2+Z^2=1$, that is, the three-dimensional state space is naturally reduced to the unit sphere.

$$X=\frac{2\sqrt{\mathcal{J}_1\mathcal{J}_2}}{\mathcal{J}_{\text{sum}}}\cos\delta,Y=\frac{2\sqrt{\mathcal{J}_1\mathcal{J}_2}}{\mathcal{J}_{\text{sum}}}\sin\delta,Z=\frac{\mathcal{J}_1-\mathcal{J}_2}{\mathcal{J}_{\text{sum}}}\tag{30}$$

The inverse transformation is

$$\mathcal{J}_1=\frac{\mathcal{J}_{\text{sum}}}{2}(1+Z),\ \mathcal{J}_2=\frac{\mathcal{J}_{\text{sum}}}{2}(1-Z),\delta=\text{atan2}(Y,X)\tag{31}$$

From the steady-state active-power expression (15), the line transmission power is $P_{12}=|u_1||u_2|\sin\delta/(\omega L)$. When $|u_1|=|u_2|=V$ ($V$ is the rated voltage) and $\sin\delta=1$, the

maximum transmission power reaches $P_{\max} = V^2/(\omega L)$. Combining this with the total action $\mathcal{J}_{\text{sum}} = \mathcal{J}_1 + \mathcal{J}_2 = V^2/(\omega^3 L)$, we obtain $P_{\max} = \omega^2 \mathcal{J}_{\text{sum}}$. It is easy to verify the identity shown in (32) that the Hamiltonian function $H$ depends only on $X$.

$$H = \omega \mathcal{J}_{\text{sum}} (1 - X) = \frac{P_{\max}}{\omega} (1 - X) \tag{32}$$

We plot the Hamiltonian function (32) as contour lines on the Bloch sphere, as shown in Fig. 2. The spherical surface in the above figure is unfolded onto a two-dimensional plane using latitude-longitude coordinates, as shown in Fig. 3.

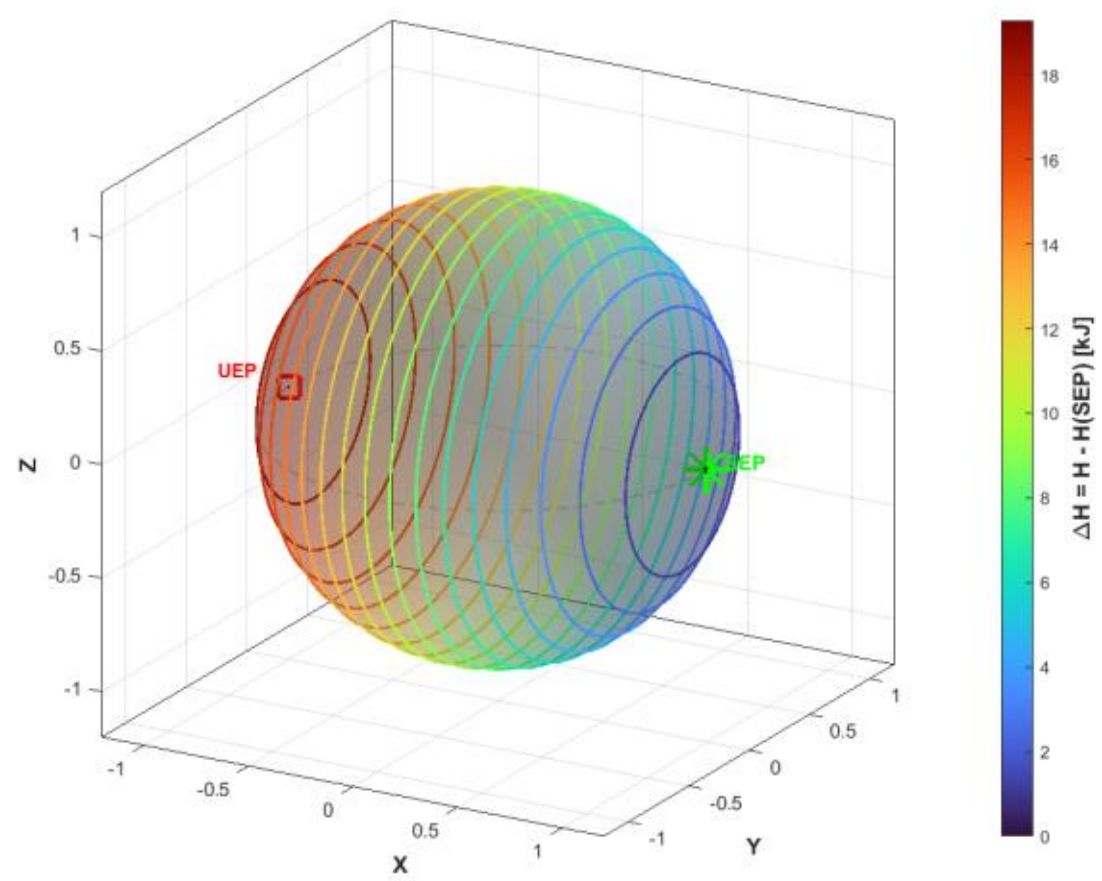

**Fig. 2.** Contour of Hamiltonian function on the Bloch sphere.

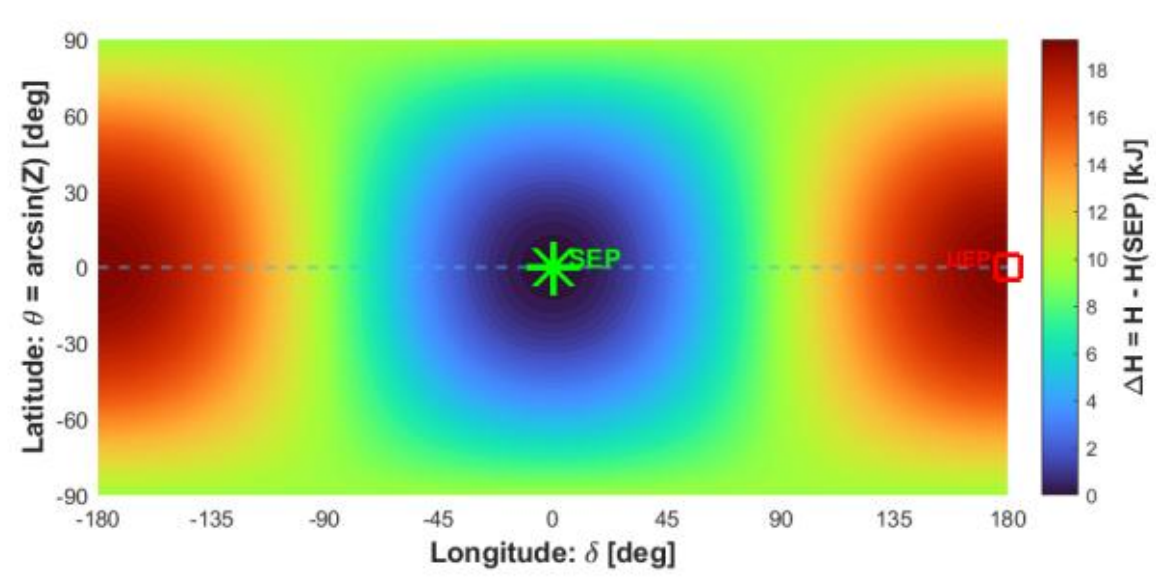

**Fig. 3.** Energy landscape in latitude-longitude coordinates.

In the above energy landscape, two equilibrium points — SEP(Stable Equilibrium Point) and UEP(Unstable Equilibrium Point) — can be observed at 0° and 180° on the equator, respectively, which is consistent with those given by the traditional EAC (Equal-Area Criterion).

## V Modification of the Energy Landscape by Active-Power Control

The modifications to energy landscape caused by active-power *P*-*δ* control and reactive-power *Q*-*V* control should be taken into account. Firstly, we consider the active-power *P*-*δ* control, for which the energy function becomes

$$H = \omega \mathcal{J}_1 + \omega \mathcal{J}_2 - 2\omega \sqrt{\mathcal{J}_1}\sqrt{\mathcal{J}_2} \cos\delta - \frac{P_{ref}\delta}{\omega} \tag{33}$$

where the last term $-P_{ref}\delta/\omega$ represents the correction to the energy function due to active-power *P*–*δ* control.

When active-power control reaches steady state, we have

$$\frac{\partial H}{\partial \delta} = 2\omega \sqrt{\mathcal{J}_1}\sqrt{\mathcal{J}_2} \sin\delta - \frac{P_{ref}}{\omega} = 0 \tag{34}$$

According to (15), we have

$$P_{ref} = 2\omega^2 \sqrt{\mathcal{J}_1}\sqrt{\mathcal{J}_2} \sin\delta = P_{1_\text{steady}} \tag{35}$$

Therefore, according to (33), the position of the SEP can be changed by adjusting $P_{ref}$. For example, with $P_{ref} = 2\text{MW}$, the energy landscape of (33) is shown in Fig. 4.

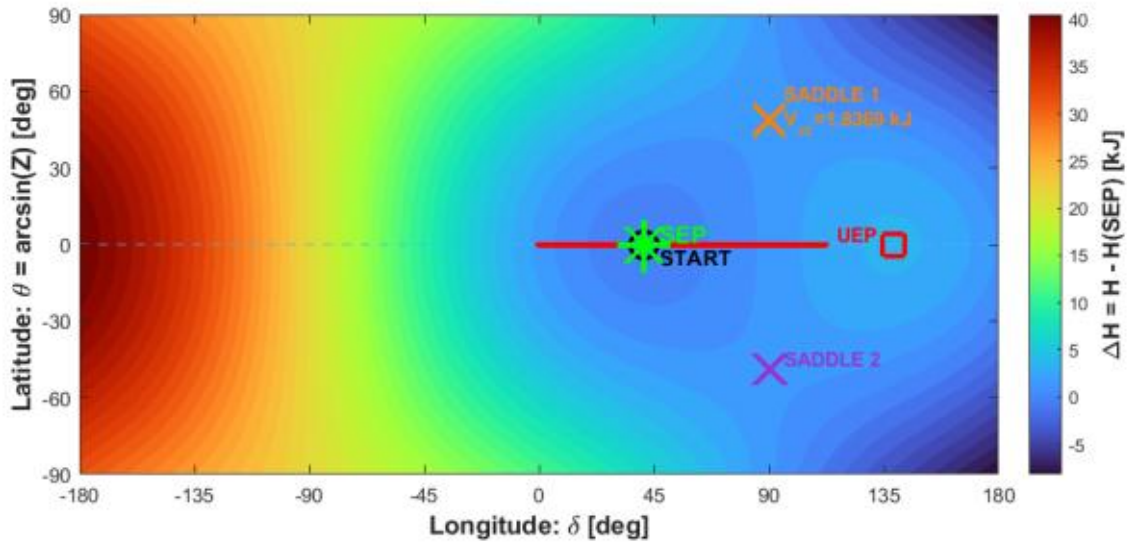

**Fig. 4.** Energy landscape when $P_{\text{ref}}$=2 MW.

In Fig. 4, the energy discontinuity between the 180° and −180° meridians arises from the term $-P_{ref}\delta/\omega$, which differs by $2\pi P_{ref}/\omega$ at $\delta = \pi$ and $\delta = -\pi$. On the equator, the SEP moves from 0° to 41.29° (−2.189kJ), while the UEP moves from 180° to 138.71° (1.485kJ). Their energy difference, i.e., the global critical energy, is 3.67kJ. Meanwhile, two saddle points symmetric about the equator appear at high latitudes. However, because traditional power system models usually assume equal voltage magnitudes at both ends (*Z*=0), the system state is constrained to the equator, and these high-latitude saddle points do not lead to instability. For example, when the VSC (Voltage Source Converter) at node 1 operates in VSM (Virtual Synchronous Machine) mode[28], the active power control equation is given by (36)[29].

$$M\ddot{\theta}_1 + D\dot{\theta}_1 = K_p \left(P_{ref} - P_1\right) \tag{36}$$

where $M = 0.1814\text{kJ}\cdot\text{s}^2/\text{rad}^2$ is the virtual inertia, $D = 0.1\text{kJ}\cdot\text{s/rad}^2$ is the virtual damping, and $K_p = 1/\omega$ is the proportional coefficient.

The VSC at node 2 is operated to keep the mean squared nodal-voltage magnitude of the transmission line invariant, the system is initially in steady state at the SEP (41.29°). Assume the system is subjected to a disturbance and thereby acquires an initial kinetic energy of 3.58 kJ. Taking the SEP as the initial state, a time-domain simulation is carried out; the resulting state trajectory on the Bloch sphere is shown as the red curve in Fig. 4, and the time evolution of this trajectory is plotted in Fig. 5. Although the system's initial energy of 3.58 kJ exceeds the critical energy defined by the saddle points, the system energy remains below the critical energy of 3.67 kJ defined by the UEP on the equator according to the traditional equal-area criterion, because the system is confined to motion along the equator. Consequently, first-swing instability does not occur.

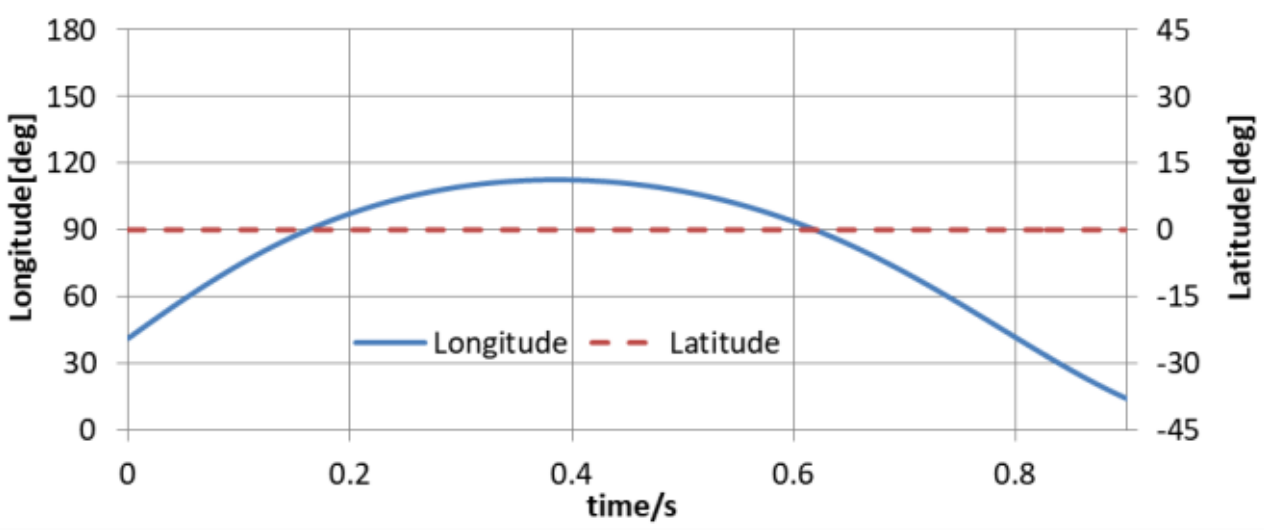

**Fig. 5.** Time-domain curves of first-swing stability on the equator.

## VI Modification of the Energy Landscape by Reactive-Power Control

We consider the energy function of the reactive-power control loop (37), which reaches its minimum energy state when $Q_{ref}-Q_1=0$. In this paper, the proportional coefficient $K_{\rm q}=0.3{\rm kJ}/\left({\rm MVar}^2\right)$.

$$U_Q=\frac{1}{2}K_{\rm q}\left(Q_{ref}-Q_1\right)^2 \tag{37}$$

Considering with (18), we have

$$\begin{aligned}Q_1&=\frac{1}{L}\left[\omega\left|\Psi_1\right|^2-\omega\left|\Psi_1\right|\left|\Psi_2\right|\cos\delta\right]\\&=2\omega^2\left[\mathcal{J}_1-\sqrt{\mathcal{J}_1\mathcal{J}_2}\cos\delta\right]\end{aligned} \tag{38}$$

$$\sqrt{\mathcal{J}_1\mathcal{J}_2}=\frac{\mathcal{J}_{\rm sum}}{2}\sqrt{1-Z^2}=\frac{\mathcal{J}_{\rm sum}}{2}\sqrt{X^2+Y^2} \tag{39}$$

Substituting (39), $\mathcal{J}_1=\frac{\mathcal{J}_{\rm sum}}{2}(1+Z)$, and $\omega^2\mathcal{J}_{\rm sum}=P_{\rm max}$ into (38), we obtain

$$Q_1(Z)=P_{\rm max}\left[(1+Z)-\sqrt{1-Z^2}\cos\delta\right] \tag{40}$$

Substituting (39) into (35), and considering $\omega^2\mathcal{J}_{\rm sum}=P_{\rm max}$, we obtain (41).

$$\cos\delta=\sqrt{1-\frac{P_{ref}^2}{P_{\rm max}^2\left(1-Z^2\right)}} \tag{41}$$

Substituting (41) into (40), we have

$$Q_1\left(Z,P_{ref}\right)=P_{\rm max}\left[(1+Z)-\sqrt{1-Z^2-\left(\frac{P_{ref}}{P_{\rm max}}\right)^2}\right] \tag{42}$$

where $Q\left(Z,P_{ref}\right)$ is a function of $Z$ and $P_{ref}$, it implicitly defines the function $Z=\left(Q,P_{ref}\right)$. With the given $Q_{ref}$ and $P_{ref}$, $Z_{ref}=Z\left(Q_{ref},P_{ref}\right)$ is accordingly determined. Hence, the reactive-power control loop energy function term (37) can be rewritten as (43).

$$U_Q=\frac{1}{2}K_q\left(Q_{ref}-Q_1\left(Z,P_{ref}\right)\right)^2 \tag{43}$$

The total energy function incorporating both active-power and reactive-power controls, is given by

$$\begin{aligned}H&=\omega\mathcal{J}_1+\omega\mathcal{J}_2-2\omega\sqrt{\mathcal{J}_1}\sqrt{\mathcal{J}_2}\cos\delta\\&-\frac{P_{ref}\delta}{\omega}+\frac{1}{2}K_q\left(Q_{ref}-Q_1\left(Z,P_{ref}\right)\right)^2\end{aligned} \tag{44}$$

For a grid-forming converter with *Q*-*V* control, by analogy to the second-order control equation (36) of the active-power channel, the corresponding control equation can generally be written as (45).

$$M_q\ddot{V}_1+D_q\dot{V}_1=k_q\left(Q_{ref}-Q_1\right) \tag{45}$$

where $M_q$ and $D_q$ are the virtual inertia and virtual damping of the reactive power channel, the proportional coefficient $k_q=K_q\,\partial Q_1/\partial V_1$.

Since node 2 keeps the mean-squared nodal voltage invariant, the latitude coordinate obeys $Z=2V_1^2/V_{\rm sum}^2-1$, $V_{\rm sum}^2=V_1^2+V_2^2$ and the inertial *Q*-*V* loop maps to the latitudinal dynamics

$$M_z\ddot{Z}+D_z\dot{Z}=-\frac{\partial H}{\partial Z} \tag{46}$$

where $M_z=M_q\left(\frac{V_{\rm sum}^2}{4V_1}\right)^2$, $D_z=D_q\left(\frac{V_{\rm sum}^2}{4V_1}\right)^2$. The energy function $H$ is given by (44). In this paper $M_z=0.05{\rm kJ}\cdot{\rm s}^2$, $D_z=0.15{\rm kJ}\cdot{\rm s}$.

Note that $Q_1\left(Z,P_{ref}\right)$ in (44) is the steady-state *Q*-*V* characteristic at $P=P_{ref}$, following the direct-method tradition[4]-[6], the energy landscape is constructed from steady-state control characteristics, which is exact at all equilibria and saddle points. Taking the reactive-power control coefficient $K_Q$=0.3kJ/MVar$^2$, the energy landscape of (44) is shown in Fig. 6.

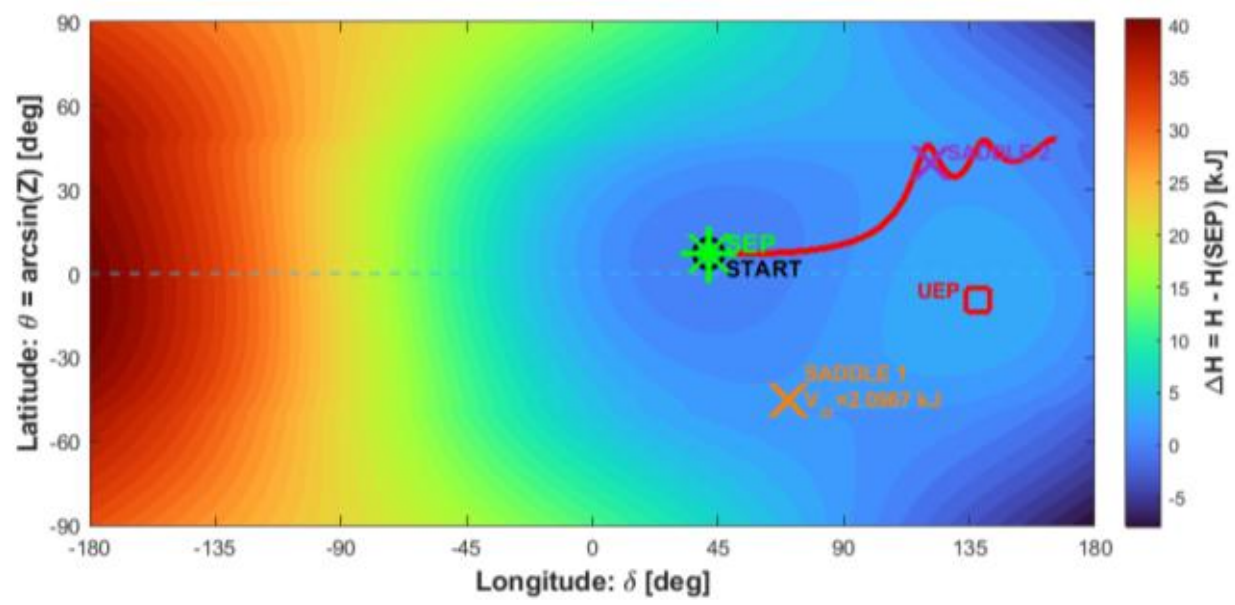

**Fig. 6.** Energy landscape when $P_{\rm ref}$= 2MW, $Q_{\rm ref}$=2 MVar.

Comparing Fig. 6 with Fig. 4, the SEP begins to shift toward high latitudes due to the introduction of the reactive-power control loop. Assuming the system is in steady-state operation (at the SEP), the same initial kinetic energy (3.58 kJ) is applied to the system, and a time-domain simulation is conducted starting from the SEP. The resulting state trajectory on the Bloch sphere is shown as the red curve in Fig. 6, and the time evolution of this trajectory is plotted in Fig. 7.

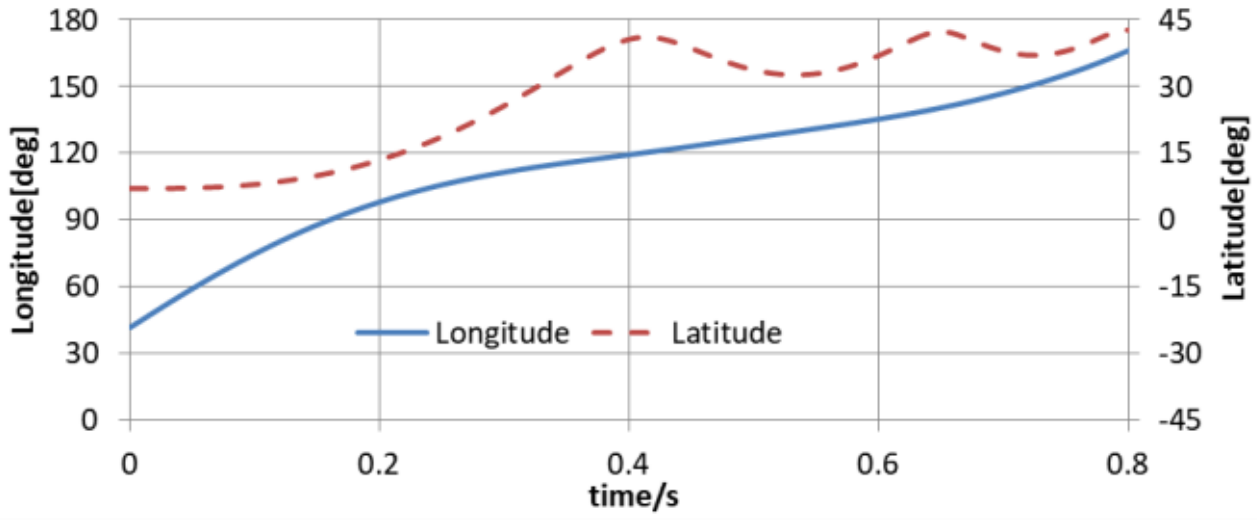

**Fig. 7.** Time-domain curves showing instability beyond the saddle point.

It can be seen that the introduction of *Q-V* loop shifts the SEP from the equator (voltage equilibrium) toward high latitudes. More importantly, the *Q-V* loop enables the system state to move across latitudes. Although the initial kinetic energy of 3.58 kJ remains below the 3.84 kJ critical energy required to cross the UEP, it already exceeds the 2.10 kJ and 2.52 kJ critical energy required to cross the saddle point 1 and 2. Consequently, in the case of insufficient damping, the system state may escape through the saddle point.

**Remark I (Validity of the equal-area criterion).** The above comparison reveals that, once the *Q*–*V* control loop is introduced, the SEP shifts away from the equator, and the stability boundary is governed by the lower-energy saddle points rather than the equatorial UEP. Conversely, under *P*–*δ* control alone with constant nodal-voltage magnitudes, the state is confined to the equator ($Z$=0), reactive-power balance is automatically maintained, and the classical equal-area criterion remains valid.

## VII Networks with n Nodes and m Branches

This section discusses extending the action-angle variable model from a two-node single-branch system to n-node m-branch networks. Consider branch $k$ connected between nodes $i$ and $j$, with branch inductance $L_k$, its magnetic energy is given by

$$H_k = \frac{1}{2L_k}\left|\Psi_i - \Psi_j\right|^2, \text{ where } 1 \le k \le m, 1 \le i, j \le n \tag{47}$$

Expanding it yields

$$H_k = \frac{1}{2L_k}\left[\left|\Psi_i\right|^2 + \left|\Psi_j\right|^2 - 2\left|\Psi_i\right|\left|\Psi_j\right|\cos\left(\theta_i - \theta_j\right)\right] \tag{48}$$

Define the nodal action $\mathcal{J}_i = \left|\Psi_i\right|^2 / \left(2\omega L_0\right)$, where $L_0$ is a reference inductance introduced to ensure dimensional correctness of the action $\mathcal{J}_k$. It is convenient to let $L_0 = 1\text{H}$. Then, (48) becomes

$$H_k = \frac{\omega L_0}{L_k}\left[\mathcal{J}_i + \mathcal{J}_j - 2\sqrt{\mathcal{J}_i \mathcal{J}_j}\cos\left(\theta_i - \theta_j\right)\right] \tag{49}$$

Summing over all $m$ branches,

$$\begin{aligned} H = \sum_{k=1}^{m} H_k &= \omega L_0 \sum_{k=1}^{m} \frac{1}{L_k}\left(\mathcal{J}_i + \mathcal{J}_j\right) \\ &\quad - 2\omega L_0 \sum_{k=1}^{m} \frac{1}{L_k}\sqrt{\mathcal{J}_i \mathcal{J}_j}\cos\left(\theta_i - \theta_j\right) \end{aligned} \tag{50}$$

Rearranging the first term of (50) into nodal form while keeping the second term in branch form yields

$$\begin{aligned} H &= \omega L_0 \sum_{i=1}^{n} \mathcal{J}_i \left( \sum_{k \in \mathcal{N}_i} \frac{1}{L_k} \right) \\ &\quad - 2\omega L_0 \sum_{k=1}^{m} \frac{1}{L_k}\sqrt{\mathcal{J}_i \mathcal{J}_j}\cos\left(\theta_i - \theta_j\right) \end{aligned} \tag{51}$$

where $\mathcal{N}_i$ denotes the set of all branches connected to node $i$.

The complete PHS equation is given by

$$\dot{x} = J\nabla H\left(x\right) + g\left(x\right)u \tag{52}$$

where the state vector is $x = \left(\mathcal{J}_1, \theta_1, \ldots, \mathcal{J}_n, \theta_n\right)^{\mathrm{T}}$.

The interconnection matrix is $J_{2n\times 2n} = \begin{bmatrix} J_1 & 0 & \cdots & 0 \\ 0 & J_2 & \cdots & 0 \\ \vdots & \vdots & \ddots & \vdots \\ 0 & 0 & \cdots & J_n \end{bmatrix}$, $J_i = \begin{bmatrix} 0 & 1 \\ -1 & 0 \end{bmatrix}$ corresponds to the symplectic block for node $i$.

Energy gradients are

$$\begin{aligned} \frac{\partial H}{\partial \theta_i} &= 2\omega L_0 \sum_{k\in\mathcal{N}_i} \frac{1}{L_k}\sqrt{\mathcal{J}_i\mathcal{J}_j}\sin\left(\theta_i - \theta_j\right) = \frac{P_{i,network}}{\omega} \\ \frac{\partial H}{\partial \mathcal{J}_i} &= \omega L_0 \sum_{k\in\mathcal{N}_i} \frac{1}{L_k} - \omega L_0 \sum_{k\in\mathcal{N}_i} \frac{1}{L_k}\sqrt{\frac{\mathcal{J}_j}{\mathcal{J}_i}}\cos\left(\theta_i - \theta_j\right) \\ &= \frac{\omega L_0}{\sqrt{\mathcal{J}_i}} \sum_{k\in\mathcal{N}_i} \frac{1}{L_k}\left(\sqrt{\mathcal{J}_i} - \sqrt{\mathcal{J}_j}\cos\left(\theta_i - \theta_j\right)\right) \\ &= \frac{Q_{i,network}}{2\omega\mathcal{J}_i} \end{aligned}$$

where $P_{i,network}$ and $Q_{i,network}$ represent the total active and reactive power injected into the network from node $i$, respectively.

The input vector $u = \left(u_{1x}, u_{1y}, \ldots, u_{nx}, u_{ny}\right)^{\mathrm{T}}$. The input matrix $g\left(x\right) = \begin{bmatrix} g_1 & 0 & \cdots \\ 0 & g_2 & \cdots \\ \vdots & \vdots & \ddots \end{bmatrix}$, where each block $g_i = \begin{bmatrix} \gamma_i\cos\theta_i & \gamma_i\sin\theta_i \\ -\eta_i\sin\theta_i & \eta_i\cos\theta_i \end{bmatrix}$, $\gamma_i = \sqrt{\frac{2\mathcal{J}_i}{\omega L_0}}$, $\eta_i = \frac{\gamma_i}{2\mathcal{J}_i}$.

It can be seen that the action-angle variable PHS model of the *n*-node *m*-branch power networks is a natural extension of the two-node system, reflecting the universal property of the adiabatic equation of the power grid under the action-angle variable framework. The adiabatic invariant is $\mathcal{J}_{\text{sum}} = \sum_{i=1}^{n} \mathcal{J}_i$.

Conserving the total action is tantamount to keeping the mean squared nodal-voltage magnitude of the AC network invariant.

## VIII CONCLUSIONS

This paper addresses the problems of time-scale coupling and fragmented energy language faced by the traditional power system analysis framework under high penetration of grid-forming converters, and proposes the canonical regularized modeling method for AC power grids based on action-angle variables. The main conclusions are as follows:

**(1) From Rigidity to Adiabaticity: Symplectic Regularization of Power Network Dynamics.** Exploiting the uniform circular motion of rotating magnetic fields in the synchronous frame, we construct action-angle coordinates from three-phase AC line inductance without shunt capacitors, embedding power-flow algebraic constraints into the PHS framework. The total action is Noether conserved in the undamped case and becomes an adiabatic invariant under slow parameter variations, transforming the grid's rigid algebraic constraints into an energy flow on a slowly varying symplectic manifold.

**(2) Visualization of Energy Landscape and Latitudinal Instability.** By introducing Bloch-sphere coordinates, this paper maps the magnetic energy of transmission lines onto the unit sphere, revealing a latitudinal instability channel that the traditional equal-area criterion fails to capture. When reactive-power voltage control is introduced by grid-forming converters, the stable equilibrium point deviates from the equator, and the stability-region boundary is determined by saddle points on the energy landscape rather than by the unstable equilibrium points on the equator.

**(3) Unified Criterion and Controllable Landscape.** The energy landscape unifies *P*–*δ* and *Q*–*V* stability into a single framework, providing an analytical platform for grid-forming converter control. The controller structure and parameters directly govern the geometric deformation of the energy landscape, laying the theoretical foundation for subsequent energy-shaping-based stabilization design.

Future work will focus on developing closed-loop energy-shaping control strategies based on the proposed canonical regularization framework, with particular emphasis on passivity-based transient stabilization of grid-forming converters in multi-node networks.

## APPENDIX A

**Problem:** In the *xy* coordinates, the integral relationship between the nodal flux linkage and nodal voltage is given by (A.1).

$$\begin{bmatrix}\dot{\Psi}_x\\ \dot{\Psi}_y\end{bmatrix}=\begin{bmatrix}0 & \omega\\ -\omega & 0\end{bmatrix}\begin{bmatrix}\Psi_x\\ \Psi_y\end{bmatrix}+\begin{bmatrix}u_x\\ u_y\end{bmatrix} \tag{A.1}$$

By introducing $\mathcal{J}=|\Psi|^2/(2\omega L)$ and $\theta=\text{angle}(\Psi)$, where $\Psi=\Psi_x+j\Psi_y$, prove that (A.2) holds.

$$\begin{cases}\dot{\mathcal{J}}=0+\sqrt{\dfrac{2\mathcal{J}}{\omega L}}\left(u_x\cos\theta+u_y\sin\theta\right)\\ \dot{\theta}=-\omega+\dfrac{-u_x\sin\theta+u_y\cos\theta}{\sqrt{2\omega L\mathcal{J}}}\end{cases} \tag{A.2}$$

**Proof: A.1 Prove the first equation of (A.2):**

Differentiate $\mathcal{J}=\dfrac{|\Psi|^2}{2\omega L}=\dfrac{\Psi\cdot\Psi^*}{2\omega L}$ with respect to time, where $\Psi^*$ denotes the complex conjugate of $\Psi$.

$$\begin{aligned}\dot{\mathcal{J}}&=\frac{\dot{\Psi}\cdot\Psi^*+\Psi\cdot\dot{\Psi}^*}{2\omega L}\\ &=\frac{\dot{\Psi}\cdot\Psi^*+\left(\dot{\Psi}\cdot\Psi^*\right)^*}{2\omega L}=\frac{\text{Real}\left(\dot{\Psi}\cdot\Psi^*\right)}{\omega L}\end{aligned} \tag{A.3}$$

Considering $\dot{\Psi}=-j\omega\Psi+u$,

$$\dot{\Psi}\cdot\Psi^*=\left(-j\omega\Psi+u\right)\Psi^*=-j\omega|\Psi|^2+u\Psi^* \tag{A.4}$$

Taking the real part of the above equation,

$$\text{Re}\left(\dot{\Psi}\cdot\Psi^*\right)=\text{Re}\left(u\Psi^*\right) \tag{A.5}$$

Therefore,

$$\dot{\mathcal{J}}=\frac{\text{Re}\left(u\Psi^*\right)}{\omega L} \tag{A.6}$$

Considering $\Psi=\sqrt{2\omega L\mathcal{J}}e^{j\theta}$, $\Psi^*=\sqrt{2\omega L\mathcal{J}}e^{-j\theta}$ and $u=u_x+ju_y$.

$$\begin{aligned}u\Psi^*&=\left(u_x+ju_y\right)\sqrt{2\omega L\mathcal{J}}e^{-j\theta}\\ &=\sqrt{2\omega L\mathcal{J}}\left(u_x+ju_y\right)\left(\cos\theta-j\sin\theta\right)\end{aligned} \tag{A.7}$$

Taking the real part of the above equation yields

$$\text{Re}\left(u\Psi^*\right)=\sqrt{2\omega L\mathcal{J}}\left(u_x\cos\theta+u_y\sin\theta\right) \tag{A.8}$$

Substituting (A.8) into (A.6), we obtain

$$\dot{\mathcal{J}}=\sqrt{\frac{2\mathcal{J}}{\omega L}}\left(u_x\cos\theta+u_y\sin\theta\right) \tag{A.9}$$

Thus, the first equation of (A.2) is proved.

**A.2 Prove the second equation of (A.2):**

First, consider the logarithmic derivative. For complex function $\Psi=|\Psi|e^{j\theta}$, its logarithmic derivative is

$$\frac{\dot{\Psi}}{\Psi}=\frac{d}{dt}\ln\Psi=\frac{d}{dt}\left(\ln|\Psi|+j\theta\right)=\frac{|\dot{\Psi}|}{|\Psi|}+j\dot{\theta} \tag{A.10}$$

For the logarithmic derivative of a complex function, the real part represents the relative magnitude change rate and the imaginary part represents the angular change rate. From (A.10), we directly obtain

$$\dot{\theta}=\text{Im}\left(\frac{\dot{\Psi}}{\Psi}\right) \tag{A.11}$$

Rewriting the above equation as

$$\dot{\theta}=\text{Im}\left(\frac{\dot{\Psi}}{\Psi}\right)=\text{Im}\left(\frac{\dot{\Psi}\Psi^*}{\Psi\Psi^*}\right)=\frac{\text{Im}\left(\dot{\Psi}\Psi^*\right)}{|\Psi|^2} \tag{A.12}$$

From (A.4), taking the imaginary part yields

$$\operatorname{Im}\left(\dot{\Psi}\Psi^*\right) = -\omega|\Psi|^2 + \operatorname{Im}\left(u\Psi^*\right) \tag{A.13}$$

From (A.7), taking the imaginary part yields

$$\operatorname{Im}\left(u\Psi^*\right) = \sqrt{2\omega L\mathcal{J}}\left(-\sin\theta\cdot u_x + \cos\theta\cdot u_y\right) \tag{A.14}$$

Therefore,

$$\begin{aligned}\dot{\theta} &= \frac{-\omega|\Psi|^2 + \sqrt{2\omega L\mathcal{J}}\left(-\sin\theta\cdot u_x + \cos\theta\cdot u_y\right)}{|\Psi|^2} \\ &= -\omega + \frac{1}{\sqrt{2\omega L\mathcal{J}}}\left(-\sin\theta\cdot u_x + \cos\theta\cdot u_y\right)\end{aligned} \tag{A.15}$$

Thus, the second equation of (A.2) is proved.

## APPENDIX B

**Problem:** Taking the two-node single-branch system as an example, prove that equation (B.1) in the rotating *xy* coordinates is equivalent to the conventional power flow equations when dynamic processes are neglected.

$$\frac{d}{dt}\begin{bmatrix}\Psi_{1x}\\ \Psi_{1y}\\ \Psi_{2x}\\ \Psi_{2y}\end{bmatrix} = \begin{bmatrix}0 & \omega & 0 & 0\\ -\omega & 0 & 0 & 0\\ 0 & 0 & 0 & \omega\\ 0 & 0 & -\omega & 0\end{bmatrix}\begin{bmatrix}\Psi_{1x}\\ \Psi_{1y}\\ \Psi_{2x}\\ \Psi_{2y}\end{bmatrix} + \begin{bmatrix}u_{1x}\\ u_{1y}\\ u_{2x}\\ u_{2y}\end{bmatrix} \tag{B.1}$$

**Proof:** Introducing complex variables $\Psi_i = \Psi_{ix} + j\Psi_{iy}$ and $u_i = u_{ix} + ju_{iy}$, rewrite (B.1) in the complex domain as

$$\frac{d}{dt}\begin{bmatrix}\Psi_1\\ \Psi_2\end{bmatrix} = -j\omega\begin{bmatrix}\Psi_1\\ \Psi_2\end{bmatrix} + \begin{bmatrix}u_1\\ u_2\end{bmatrix} \tag{B.2}$$

Neglecting the dynamic processes, we have

$$\begin{bmatrix}u_1\\ u_2\end{bmatrix} = j\omega\begin{bmatrix}\Psi_1\\ \Psi_2\end{bmatrix} \tag{B.3}$$

Multiplying both sides of (B.3) by $\frac{1}{j\omega L}\begin{bmatrix}1 & -1\\ -1 & 1\end{bmatrix}$ yields

$$\frac{1}{j\omega L}\begin{bmatrix}1 & -1\\ -1 & 1\end{bmatrix}\begin{bmatrix}u_1\\ u_2\end{bmatrix} = \frac{1}{L}\begin{bmatrix}1 & -1\\ -1 & 1\end{bmatrix}\begin{bmatrix}\Psi_1\\ \Psi_2\end{bmatrix} = \begin{bmatrix}I_1\\ I_2\end{bmatrix} \tag{B.4}$$

Equation (B.4) gives the conventional nodal admittance equation, as shown in (B.5).

$$G\begin{bmatrix}u_1\\ u_2\end{bmatrix} = \begin{bmatrix}I_1\\ I_2\end{bmatrix}, \text{ where } G = \frac{1}{j\omega L}\begin{bmatrix}1 & -1\\ -1 & 1\end{bmatrix} \tag{B.5}$$

The nodal injection power is given by (B.6).

$$P_1 + jQ_1 = u_1\cdot I_1^*,\ P_2 + jQ_2 = u_2\cdot I_2^* \tag{B.6}$$

Equations (B.5) and (B.6) constitute the conventional power flow algebraic equations together.

## REFERENCES

[1] P. Kundur, *Power System Stability and Control*. New York, NY, USA: McGraw-Hill, 1994.

[2] P. W. Sauer and M. A. Pai, *Power System Dynamics and Stability*. Upper Saddle River, NJ, USA: Prentice-Hall, 1998.

[3] F. Milano, *Power System Modelling and Scripting*. Berlin, Germany: Springer, 2010.

[4] H. D. Chiang, F. F. Wu, and P. P. Varaiya, "Foundations of direct methods for power system transient stability analysis," in *IEEE Transactions on Circuits and Systems*, vol. 34, no. 2, pp. 160-173, February 1987, doi: 10.1109/TCS.1987.1086115.

[5] H. D. Chiang, F. F. Wu, and P. P. Varaiya, "Foundations of the potential energy boundary surface method for power system transient stability analysis," in *IEEE Transactions on Circuits and Systems*, vol. 35, no. 6, pp. 712-728, June 1988, doi: 10.1109/31.1808.

[6] H. D. Chiang, F. F. Wu, and P. P. Varaiya, "A BCU method for direct analysis of power system transient stability," in *IEEE Transactions on Power Systems*, vol. 9, no. 3, pp. 1194-1208, Aug. 1994, doi: 10.1109/59.336079.

[7] U. Markovic, O. Stanojev, P. Aristidou, and G. Hug, "Understanding small-signal stability of low-inertia systems," in *IEEE Transactions on Power Systems*, vol. 36, no. 5, pp. 3997-4017, Sept. 2021, doi: 10.1109/TPWRS.2021.3061434.

[8] X. Jiang, C. M. Lagoa, and Y. Li, "Power system electromagnetic transient stability: An analysis based on convergent Hamiltonian," arXiv preprint arXiv:2502.09695, 2025.

[9] X. Chen *et al.*, "Transient Stability Analysis and Enhancement of Grid-Forming Converters: A Comprehensive Review," *Electronics*, vol. 14, no. 4, p. 645, 2025, doi: 10.3390/electronics14040645.

[10] A. J. van der Schaft and B. M. Maschke, "The Hamiltonian formulation of energy conserving physical systems with external ports," *Arch. Elek. Übertragungstech.*, vol. 49, pp. 362–371, 1995, doi: 10.1016/0308-5961(95)90007-1.

[11] A. J. van der Schaft, *L2-Gain and Passivity Techniques in Nonlinear Control*, 3rd ed. Cham, Switzerland: Springer, 2017.

[12] T. Stegink, C. De Persis and A. van der Schaft, "A Unifying Energy-Based Approach to Stability of Power Grids With Market Dynamics," in *IEEE Transactions on Automatic Control*, vol. 62, no. 6, pp. 2612-2622, June 2017, doi: 10.1109/TAC.2016.2613901.

[13] S. Fiaz, D. Zonetti, R. Ortega, J.M.A. Scherpen, A.J. van der Schaft, "A port-Hamiltonian approach to power network modeling and analysis", in *European Journal of Control*, Volume 19, Issue 6, 2013, Pages 477-485, doi: 10.1016/j.ejcon.2013.09.002.

[14] Q.-C. Zhong and M. Stefanello, "Generic Modeling and Control Framework for Power Systems Dominated by Power Converters Connected Through a Passive Transmission and Distribution Grid," *CSEE J. Power Energy Syst.*, vol. 10, no. 1, pp. 292-301, January 2024, doi: 10.17775/CSEEJPES.2023.06400.

[15] Papastavridis, J. G., *Analytical Mechanics: A Comprehensive Treatise on the Dynamics of Constrained Systems*, corrected reprint of the 2002 original. Hackensack, NJ, USA: World Scientific, 2014.

[16] L. D. Landau and E. M. Lifshitz, *Mechanics*, 3rd ed., Course of Theoretical Physics, vol. 1. Oxford, U.K.: Butterworth-Heinemann, 1976.

[17] H. Goldstein, C. P. Poole, and J. L. Safko, *Classical Mechanics*, 3rd ed. San Francisco, CA, USA: Addison Wesley, 2002.

[18] D. Brouwer and G. M. Clemence, *Methods of Celestial Mechanics*. New York, NY, USA: Academic Press, 1961.

[19] A. Morbidelli, *Modern Celestial Mechanics: Aspects of Solar System Dynamics*. London, U.K.: Taylor & Francis, 2002.

[20] L. D. Landau and E. M. Lifshitz, *Quantum Mechanics: Non-Relativistic Theory*, 3rd ed., Course of Theoretical Physics, vol. 3. Oxford, U.K.: Butterworth-Heinemann, 1991.

[21] T. G. Northrop, *The Adiabatic Motion of Charged Particles*. New York, NY, USA: Interscience, 1963.

[22] F. Ji, L. Gao and C. Lin, "Lagrangian Modelling and Motion Stability of Synchronous Generator-based Power Systems," *CSEE J. Power Energy Syst.*, vol. 11, no. 1, pp. 13–23, Jan. 2025, doi: 10.17775/CSEEJPES.2024.00780.

[23] F. Ji, L. Gao and C. Lin, "Dynamics of three-phase AC system and VSC access problem research," *Proc. CSEE*, vol. 42, no. 6, pp. 2286–2298, 2022, doi: 10.13334/j.0258-8013.pcsee.210027.

[24] M. A. Nielsen and I. L. Chuang, *Quantum Computation and Quantum Information*. Cambridge, U.K.: Cambridge Univ. Press, 2000.

[25] F. Bloch, "Nuclear induction," *Phys. Rev.*, vol. 70, no. 7–8, pp. 460–474, 1946.

[26] L. Allen and J. H. Eberly, *Optical Resonance and Two-Level Atoms*. Mineola, NY, USA: Dover, 1987.

[27] H. Akagi, E. H. Watanabe, and M. Aredes, *Instantaneous Power Theory and Applications to Power Conditioning*. Hoboken, NJ, USA: Wiley-Interscience, 2007.

[28] Q.-C. Zhong and G. Weiss, "Synchronverters: Inverters that mimic synchronous generators," in *IEEE Transactions on Industrial*

*Electronics*, vol. 58, no. 4, pp. 1259-1267, April 2011, doi: 10.1109/TIE.2010.2048839.

[29] W. Wang, Z. Liu, J. Lou, C. Cheng, Q. Song, and Y. Shi, "Analysis of broadband oscillation mechanisms in grid-forming and grid-following converters based on virtual synchronous generator," *IET Gener. Transm. Distrib.*, vol. 19, no. 1, pp. 1-12, Feb. 2025, doi: 10.1049/gtd2.70015.